\documentclass[twocolumn,amsmath,amssymb,floatfix,superscriptaddress,prl,hyperref]{revtex4}
\usepackage[usenames,dvipsnames]{xcolor}
\usepackage{graphicx}
\usepackage{dcolumn}  
\usepackage{bm}  
\usepackage{ulem}
\usepackage{hyperref}
\usepackage{nicefrac}
\allowdisplaybreaks[2]

\begin{document}

\title{Correlating toughness and roughness in ductile fracture}

\author{L. Ponson}\email{laurent.ponson@upmc.fr}
\affiliation{ Institut Jean le Rond d'Alembert (UMR 7190),CNRS - Universit{\'e} Pierre et Marie Curie, Paris, France}
\author{A. Srivastava}
\author{S. Osovski}
\affiliation{Department of Materials Science \& Engineering,
University of North Texas, Denton, TX USA}
\author{E. Bouchaud}
\affiliation{ ESPCI, Paris Tech, Paris, France}
\author{V. Tvergaard}
\affiliation{Department of Mechanical Engineering,
The Technical University of Denmark, Lyngby, Denmark}
\author{A. Needleman}
\affiliation{Department of Materials Science \& Engineering,
University of North Texas, Denton, TX USA}

\begin{abstract}
Three dimensional calculations of ductile crack growth under mode I plane strain, small scale yielding conditions are carried out using an elastic-viscoplastic constitutive relation for a progressively cavitating plastic solid with two populations of void nucleating second phase particles. Full field solutions are obtained for three dimensional material microstructures characterized by random distributions of void nucleating particles. Crack growth resistance curves and fracture surface roughness statistics are calculated using standard procedures. The range of void nucleating particle volume fractions considered give rise to values of toughness, $J_{\mathrm IC}$, that vary by a factor of four. For all volume fractions considered, the computed fracture surfaces are self-affine over a size range of about two orders of magnitude with a roughness exponent of $0.54 \pm 0.03$. For small void nucleating particle volume fractions, the mean large particle spacing serves as a single dominant length scale. In this regime, the correlation length of the fracture surface corresponding to the cut-off of the self-affine behavior is found to be linearly related to $J_{\mathrm IC}$ thus quantitatively correlating toughness and fracture surface roughness.
\end{abstract}

\maketitle 

Thirty years ago, Mandelbrot and coworkers revealed the self-affine nature of fracture surfaces \cite{Mandelbrot}. Their hope was to relate the roughness of fracture surfaces via the exponents characterizing their scale invariance properties to the material's crack growth resistance. This hope has remained unfulfilled. Indeed, later studies showed that the value of the roughness exponent was not only independent of the material toughness but also of the material considered, as long as the failure mechanism remained the same \cite{Bouchaud9,Maloy}. Indeed, the scaling exponent measured along the propagation direction was observed to take a value $\beta^{\mathrm{brittle}} \simeq 0.5$ rather independent of the considered material for brittle failure while another value around $\beta^{\mathrm{damage}} \simeq 0.6$ was observed for damage accompanying failure \cite{Bonamy6,Ponson6}.

The universality of fracture surface roughness exponents limits the applicability of quantitative fractography based on statistical analyses for the characterization of microscopic failure mechanisms and toughness. On the other hand, it has paved the way for a unified theoretical framework based on critical transition theory to describe the failure properties of disordered materials. By interpreting the onset of material failure as a dynamic phase transition, many aspects of the behavior of cracks in disordered materials has thus been rationalized, such as the intermittent dynamics of cracks \cite{Maloy3,Bonamy5}, their scale invariant roughness \cite{Schmittbuhl4,Santucci7}, their average dynamics \cite{Koivisto,Ponson14} and their effective toughness \cite{Demery,Patinet2}. Most of these successes have been achieved in the context of brittle failure, but our understanding of the scaling properties of ductile fracture surfaces is still limited.

The process that governs the ductile fracture of structural materials at room temperature is one of nucleation, growth and coalescence of micron scale voids, and involves large plastic deformations. Quantitative models of crack growth by the progressive coalescence of voids with a crack have been available since the 1970s \cite{rice69,aravas,viggo,afek}, and calculations have provided reasonable agreement with experimental toughness measurements \cite{becker}. However, only recently has the capability been developed to calculate sufficient amounts of three dimensional ductile crack growth in heterogeneous microstructures to obtain a statistical characterization of the predicted fracture surfaces \cite{Needleman, Ponson}. This enables us to explore the microscopic mechanisms governing the fracture surface roughness as well as the relation, if any, to a material's crack growth resistance.

In this study, we capitalize on these new developments and show that the scaling properties of ductile cracks can correlate with the material's toughness. However, the relation is not with the value of the roughness exponent, but with the cut-off length of the scale invariant regime. In particular, we show that the cut-off length scale of the self-affine behavior of ductile cracks can be quantitatively related to a measure of fracture toughness. This correlation is shown in our simulations by varying one parameter of the material microstructure resulting in a family of ductile materials with a broad range of toughness.

\noindent \textit{Model formulation} \--- The connection between the roughness of ductile cracks and the material's toughness is investigated through the finite element analysis of transient three dimensional boundary value problems. A mode I small scale yielding boundary value problem is analyzed with symmetry conditions corresponding to an overall plane strain constraint. Remote displacement boundary conditions corresponding to the quasi-static linear isotropic elastic mode I crack tip stress intensity factor $K_I$ are prescribed. A finite deformation continuum mechanics formulation for a progressively cavitating solid is used. The constitutive framework is a modified Gurson constitutive relation for a progressively cavitating solid (often termed the GTN relation, see \cite{Viggo2}) with slight material rate dependence and with the plastic flow potential $\Phi$ given by
\begin{equation}
\Phi =\frac{\sigma _{e}^{2}}{\bar{\sigma}^{2}}+2q_{1}f^{*}\cosh \left(
\frac{
3q_{2}\sigma _{h}} {2\bar{\sigma}}\right) -1- \left(q_{1} f^{*}\right)
^{2}=0 
\label{pot}
\end{equation} 
where $\sigma_e$ is the Mises effective stress, $\sigma_h$ is the hydrostatic stress (positive in tension), $\bar{\sigma}$ is the material flow strength, $f^*$ is a measure of the void volume fraction and $q_1=1.25$ and $q_2=1.0$. A key feature of the constitutive relation is that the material's stress carrying capacity increases due to strain and strain rate hardening, but eventually decreases due to the nucleation and growth of micro-voids, and can vanish leading to the creation of new free surface. 

The second phase particles present in conventional structural metals are the primary source of internal cavitation at least at low temperatures. Hence we characterize the material microstructure by two populations of void nucleating second phase particles: (i) homogeneously distributed small particles and (ii) discretely modeled randomly distributed large particles. For homogeneously distributed small particles, a critical strain level following a normal distribution controlled void nucleation. Void nucleation from large particles in general depends on deformation and hydrostatic stress history. Hence, nucleations occur when $\bar{\sigma}+\sigma_h$ reaches a critical value taken here also from a normal distribution. Following this stress based criteria, the large particles nucleate voids at an early stage of the deformation history. 

The large particles are randomly located, and the maximum stress based void nucleation criteria is applied in a sphere of radius $r_0$ around their center. The elastic and plastic properties of the particles and the matrix material are identical. Only the void nucleation characteristics differ. A uniform $208 \times 64 \times 10$ mesh of $20$ node brick finite elements is used in front of the initial crack. The in-plane ($x-y$ plane) element dimension is denoted by $e_x$ which serves as a normalization length. The size and spacing of the large particles introduce characteristic lengths into the formulation. The particles radius, $r_0 = 1.5 \, e_x$, is fixed and calculations are carried out for eight volume fractions, $n$, of the large particles. For each large particle volume fraction, calculations were carried out for seven random spatial distributions. A more detailed description of the problem formulation with additional references is given in \cite{Needleman, Ponson}.

\begin{figure}
\includegraphics[width=1\columnwidth]{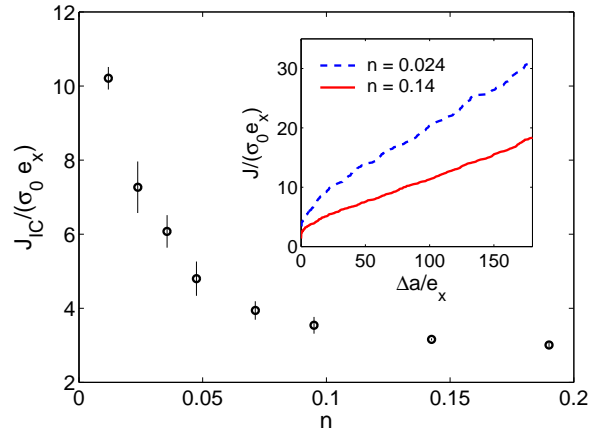}
\caption{Toughness characterization: Effect of the volume fraction $n$ of large particles on the fracture toughness $J_{\mathrm{IC}}$. Inset: J-R curves showing the material's crack growth resistance as a function of the crack extension $\Delta a$ for two large particle volume fractions.}
\label{Fig1}
\end{figure}

\noindent \textit{Toughness characterization} \--- The volume fraction of large particles is varied from $n = 0.012$ to $n = 0.19$, corresponding to mean large particle spacings of $10.6 \, e_x$ and $4.21 \, e_x$, respectively. Under plane strain conditions, the $J$-integral \cite{Rice3}, a measure of the driving force for crack propagation, is related to the applied mode I stress intensity factor, $K_I$, by
\begin{equation}
J = {K_I}^2 \left(\frac{1-\nu^2}{E} \right) 
\end{equation}
where $E = 70 \, \mathrm{GPa}$, and $\nu = 0.3$ are Young's modulus and Poisson's ratio, respectively. Figure~\ref{Fig1} shows the variation of $J_{\mathrm{IC}}$, normalized by the reference flow strength $\sigma_0=300 \,  \mathrm{MPa}$ and $e_x$, with large particle volume fraction $n$.  The error bars are calculated from realizations of large particle distributions having the same $n$. The value of $J_{\mathrm{IC}}$ characterizes the crack growth resistance and is computed using a widely used ASTM standard procedure \cite{astm}. A power law $J=C_1 \Delta a^{C_2}$ is used to fit the initial portion of the J-R curve (shown in the inset of Fig.~\ref{Fig1} for two values of $n$) and the value of $J_{\mathrm{Ic}}$ is defined as the intersection of this curve with the line $J=2 \sigma_0 \left(\Delta a - \Delta a_0\right)$, where we take $\Delta a_0/e_x = 2$. Another choice in the value of $\Delta a_0$ would change the value of $J_{\mathrm{IC}}$ but the dependence of $J_{\mathrm{IC}}$ on $n$ would remain approximately the same up to some multiplicative constant. The increase in the value of $J$ with the crack extension $\Delta a$ seen in the inset of Fig.~\ref{Fig1} is characteristic of ductile crack growth.  

From dimensional considerations alone, prediction of the $J(\Delta a)$ curve requires that the formulation contain a characteristic length. The value of $J_{\mathrm{IC}}$ increases by a factor of almost four as the volume fraction of large particles decreases or, equivalently, with an increasing mean particle spacing. Indeed, in the calculations here, the mean particle spacing serves as a microstructurally based characteristic length for the entire $J(\Delta a)$ curve as well as for $J_{\mathrm{IC}}$. 

\begin{figure}
\includegraphics[width=1\columnwidth]{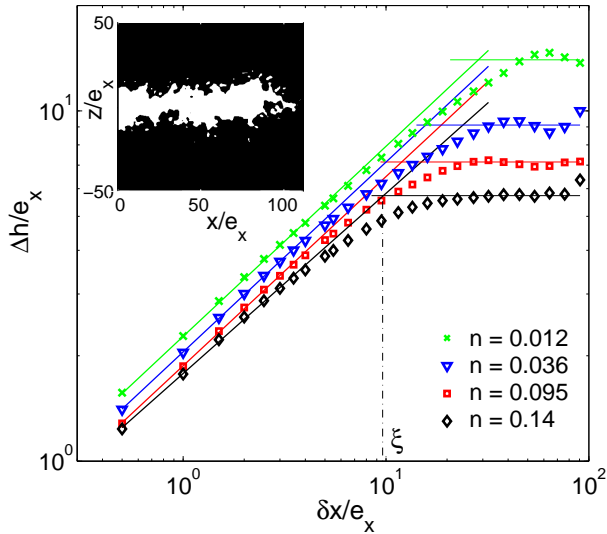}
\caption{Height-height correlation functions of the fracture surface showing the effect of the volume fraction $n$ of the large particles. Inset: Snapshot of the porosity field for a material with $n = 0.048$ showing a propagating ductile crack. The white region corresponds to a porosity larger than the threshold value $0.1$ used to define the fracture surface.}
\label{Fig2}
\end{figure}

\begin{figure}
\includegraphics[width=1\columnwidth]{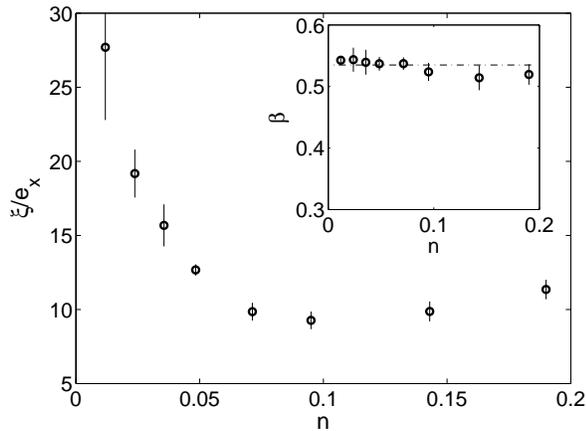}
\caption{Roughness characterization: Variation of the cut-off length $\xi$ as a function of the volume fraction $n$ of the large particles.  Inset: Variation of the roughness exponent $\beta$ with $n$.}
\label{Fig3}
\end{figure}

\noindent {\it Fracture surface characterization} \--- For each value of large particle volume fraction, $n$, fourteen statistically equivalent fracture surfaces (top and bottom surfaces) were produced, except for $n=0.012$ where ten surfaces were used. The height maps $h(x, \, y)$ were obtained from the calculated porosity field using a threshold value of void volume fraction of $0.1$, as illustrated in the inset of Fig.~\ref{Fig2} (see \cite{Ponson} for the detailed procedure). All calculations involve crack propagation over $\Delta a \simeq 180 \, e_x$, i.e. at least an order of magnitude greater than the mean large particle spacing.

The roughness is characterized using the height-height correlation function defined as
 \begin{equation}
\Delta h(\delta x)=\sqrt{<[h(x+\delta x, \, y)- h(x, \, y)]^2>_{x, \, y}}
\end{equation}
and computed on the statistically equivalent surfaces. We focus here on the correlations of height variations in the propagation direction $x$. The effect of the particle volume fraction $n$ on the fracture surface scaling is shown in Fig.~\ref{Fig2}. Regardless of the value of $n$, the correlation function follows a power law behavior at small scales and then saturates at a larger scale, indicating a self-affine behavior of the roughness up to some cut-off length $\xi$. The latter is defined at the abscissa between the power law fit of the self-affine regime and the plateau behavior at the larger scale. The first regime is characterized by the roughness exponent $\beta \simeq 0.54 \pm 0.03$ corresponding to the slope of a straight line fit in the logarithm representation of Fig.~\ref{Fig2}. This value is not affected by the large change in the particle spacing, as observed in the inset of Fig. \ref{Fig3} where the value of $\beta$ is shown as a function of the particle volume fraction $n$. This observation is in agreement with previous results obtained from similar simulations \cite{Needleman,Ponson}, and captures rather well the universal self-affine nature of ductile fracture surfaces with $\beta \simeq 0.6$ observed experimentally \cite{Ponson5}. As can be seen in  Fig.~\ref{Fig2}, both the roughness amplitude in the self-affine regime, exemplified by the vertical shift of the correlation function, and the plateau level do vary with $n$. This dependence is reflected in the cut-off length scale $\xi$ that represents the upper bound of the self-affine domain. As shown in Fig.~\ref{Fig3}, $\xi$ decreases with increasing large particle volume fraction, i.e. with decreasing mean large particle spacing.  Also, the analysis of the full statistics of the crack roughness obtained from this calculation and presented in Ref.~\cite{Ponson} shows strong deviations from the Gaussian distribution, in agreement with experimental observations \cite{Vernede}.

\begin{figure}
\includegraphics[width=1\columnwidth]{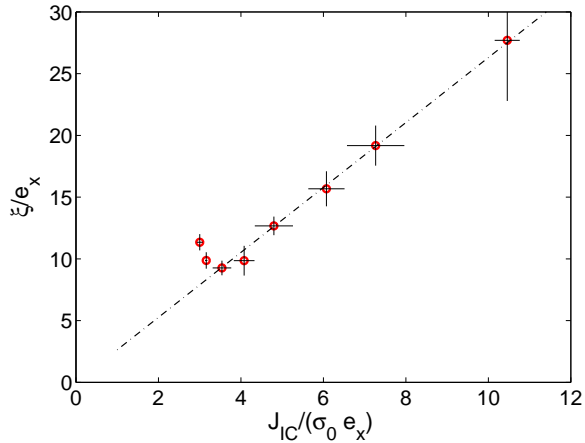}
\caption{Correlation between toughness and roughness: Variations of the cut-off length extracted from the fracture surface as a function of the normalized crack growth resistance. The straight line $\xi/e_x = \alpha J_{\mathrm{IC}}/(\sigma_0 e_x)$ going through the origin with slope $2.6$  is shown to illustrate the trend.}
\label{Fig4}
\end{figure}
\noindent {\it Toughness/roughness relationship} \--- The variation of $\xi$ with $J_{\mathrm{IC}}$ presented in Fig.~\ref{Fig4} shows a clear correlation between a measure of the ductile fracture surface roughness, $\xi$, and a measure of the material's resistance to crack growth, $J_{\mathrm{IC}}$. For brittle solids, a relation between the critical stress intensity factor $K_{IC}$ and a cutoff length is discussed in Ref.~\cite{bouch2}. The length $\xi$ can be interpreted as the typical size of the largest roughness features along the mean fracture plane. As a result, within the family of ductile solids investigated, with the exception of the two largest particle densities, the tougher the material, the rougher its fracture surface.

\noindent {\it Discussion} \--- To understand the $\xi$ versus $J_{\mathrm{IC}}$ correlation, we first examine the mechanisms that set the length scale $\xi$. Previous experimental studies on glass and mortar fracture surfaces have reported two scaling regimes $\Delta h \sim \delta x^\beta$, with $\beta^{\mathrm{damage}}  \simeq 0.6$ at small length scales $\delta x < \xi$ and $\beta^{\mathrm{brittle}} \simeq 0.5$ at larger length scales $\delta x > \xi$ \cite{Bonamy2,Morel8}. In phase-separated glass samples, two regimes were also reported, but the second regime was characterized by logarithmic correlations of height fluctuations, coinciding with $\beta^{\mathrm{brittle}} \simeq 0$ \cite{Dalmas}. These large scale behaviors could be captured quantitatively by linear elastic fracture mechanics based models of crack propagation within disordered brittle solids \cite{Ramanathan,Bonamy2}, indicating that beyond the scale $\xi$, these fracturing solids behave as a coarse-grained equivalent linear elastic medium.  For brittle and quasi-brittle solids, this suggests an interpretation of the length $\xi$ in terms of process zone size, or extension of the zone in the crack tip vicinity where linear elasticity breaks down. 

For ductile cracks, this zone corresponds to the plastic zone where plastic dissipation takes place, and can be much larger than the fracture process zone, where the actual failure processes and microcracking take place. As a consequence, the crossover length $\xi$ measured in this study between the self-affine and the plateau regimes calls for an alternative interpretation.
For a homogeneous elastic-plastic continuum in small scale yielding, the plastic zone size is linearly related to $J/\sigma_0$. When a single dominant length scale characterizes the micro-scale fracture processes, dimensional considerations require that length scale to be also linearly related to $J/\sigma_0$. For example, in models presuming an initial void near the crack tip \cite{rice69, aravas}, the critical value $J_{Ic}$ of the driving force for crack initiation is set by the distance of the void from the crack tip, leading to a linear relation between a micro-scale length characteristic of the local failure processes, the distance of the void from the crack tip, and a macro-scale length $J_{Ic}/\sigma_0$, defining the macroscopic failure properties. Also, in the model of Ref.~\cite{afek} that considers a void by void ductile failure process, the distance from the crack tip at which void nucleation occurs is found to be proportional to $J/\sigma_0$.

Figure~\ref{Fig4} shows the relation between $J_{\mathrm{IC}}$, a measure of the resistance to crack propagation and $\xi$, the correlation length of the geometrical perturbations of the fracture surface. However, both are related to the volume fraction $n$ of large particles or equivalently to the mean large particle spacing $\ell_0 \propto 1/n^{1/3}$ as seen in Figs.~\ref{Fig1} and \ref{Fig3}. In the limit of low volume fractions $n$ of large particles (corresponding to large values of $J_{\mathrm{IC}}/(\sigma_0 e_x)$), a void by void crack propagation regime, with mean large particle spacing $\ell_0$, may be the dominant mechanism. On the one hand, this would set the fracture zone size, and consequently the roughness correlation length $\xi$ to scale with $\ell_0$. On the other hand, this length scale may govern the value of the toughness, $J_{\mathrm{IC}}$, as shown in Ref.~\cite{viggo} for such a ductile crack growth mechanism in a simpler 2D geometry. Hence, as long as $\ell_0$ governs the fracture processes in the vicinity of the crack tip, both the fracture surface roughness and the overall crack growth resistance would be dominated by the single microstructural length scale $\ell_0$, leading to a linear relation between $\xi$ and $J_{\mathrm{IC}}/\sigma_0$ as seen in Fig.~\ref{Fig4}. In addition, since the plastic zone size scales with $J/\sigma_0$, $\xi$ is also linearly related to the plastic zone size.

The idealized calculation of Ref.~\cite{viggo} introduces the dimensionless parameter $C = J_{\mathrm{IC}}/\left(\sigma_0 \ell_0\right)$ that reveals which of these ductile failure mechanisms dominates. In our calculations, $C \simeq 1.0$ for $n=0.012$ and saturates to a value of $C \approx 0.68$ for $n \geq 0.071$ (a saturation in the value of $J_{\mathrm{IC}}$ and $\xi$ for $n \geq 0.071$ can also be seen in Figs.~\ref{Fig1} and \ref{Fig3}). A rough comparison with the values of $C$ obtained in Ref.~\cite{viggo} suggests that the void by void crack growth is the dominant mechanism for $n<0.071$, in agreement with the scenario proposed previously to explain the linear relation between $\xi$ and $J_{\mathrm{IC}}/(\sigma_0 e_x)$. For $n \geq 0.071$, the value of $C$ is consistent with another competing mechanism where crack growth is dominated by multiple voids (or more generally defects) interaction, accounting for the deviation from a linear relation seen at smaller values of $\xi$ in Fig.~\ref{Fig4} for greater volume fractions of large particles (or smaller mean particle spacings). A more detailed analysis of these competing mechanisms is underway. When multiple defect interactions become more prevalent, the mean particle spacing $\ell_0$ is no longer the only relevant roughness length scale and the linear relation between $\xi$ and $J_{\mathrm{IC}}/(\sigma_0 e_x)$ breaks down.

\noindent {\it Conclusion} \---Our calculations show that: (i) with a random distribution of void nucleating particles and fixed material properties, the mean particle spacing is the dominant length scale; (ii) the roughness correlation length $\xi$, corresponding to the cut-off of the self-affine behavior, reflects this length scale; (iii) $\xi$ is linearly related to $J_{\mathrm{IC}}$ as long as one length scale characterizes the microscale fracture process. These results provide an important step toward fulfilling the hope that the statistical characterization of ductile fracture surface roughness may be used for a post-mortem estimate of fracture toughness.

The financial support provided by the U.S. National Science Foundation, Grant CMMI-1200203 (AS, SO, AN), and by the European Union, ToughBridge Marie Curie Grant (LP), is gratefully acknowledged.

\end{document}